\documentclass[useAMS,usenatbib]{mn2e}

\usepackage{graphicx}
\usepackage{epsfig, floatflt}
\usepackage[varg]{txfonts}
\usepackage{longtable}

\title[Photometric iron abundance of Cepheids]
{Determination of the iron content of Cepheids from the shape of their light curves}
\author[P. Klagyivik et al.]
{P. Klagyivik$^{1}$\thanks{E-mail: klagyivik.peter@csfk.mta.hu}, L. Szabados$^{1}$,
A. Szing$^{1}$, S. Leccia$^{2}$ and N. Mowlavi$^{3,4}$\\
$^{1}$Konkoly Observatory, Research Centre for Astronomy and Earth
   Sciences, Hungarian Academy of Sciences,\\
   H-1121 Budapest XII, Konkoly Thege \'ut 15-17., Hungary\\
$^{2}$INAF -- Astronomical Observatory of Capodimonte, Salita Moiariello 16, I-80131 Napoli, Italy\\
$^{3}$ISDC Data Centre for Astrophysics, University of Geneva, Chemin d'Ecogia 16, CH-1290 Versoix, Switzerland\\
$^{4}$Department of Astronomy, University of Geneva, Chemin des Maillettes 51, CH-1290 Sauverny, Switzerland}
\begin{document}

\date{Accepted xxx. Received xxx; in original form xxx}

\pagerange{\pageref{firstpage}--\pageref{lastpage}} \pubyear{2013}

\maketitle

\label{firstpage}

\begin{abstract}

We present the study of the metallicity dependence of the Fourier amplitude 
ratios $R_{21}$ and $R_{31}$ for the light curves of short-period 
Galactic classical Cepheids in $B$, $V$, $R_C$ and $I_C$ bands. 
Based on the available photometric and spectroscopic data we determined 
the relations between the atmospheric iron abundance, [Fe/H], 
and the Fourier parameters.
Using these relations we calculated the photometric [Fe/H] values of all 
program Cepheids with an average accuracy of $\pm 0.15$ dex. 
No spectroscopic iron abundance was known before for 14 of these 
stars. These empirical results provide 
an alternate method to determine the iron abundance of classical Cepheids
too faint for spectroscopic observations. 
We also checked whether the metal-poor Cepheids of both Magellanic
Clouds follow the same relationships, and a good agreement was found.

\end{abstract}


\section{Introduction}
\label{intro}

The period-luminosity ($P$-$L$) relation of classical Cepheids is one
of the most important tools in the extragalactic distance determination.
Even though it is a well studied relationship its precision needs
improvement. There are many effects that cause a dispersion around
the ridge-line relationship. A summary of these effects is given in
\citet{szab12b}. One of these `widening' effects is the spread 
in the chemical composition of the Cepheids.

There are many observational and theoretical studies that deal
with the influence of the chemical composition on the
$P$-$L$ relationship. Theoretical computations
based on linear pulsation models (\citealt*{sandage99}; 
\citealt{Bara01}) show only a moderate influence,
while more realistic, non-linear  pulsation models 
(\citealt{Bono99}; \citealt{Caputo00}) result in a significant
dependence on the chemical composition in the sense
that metal-rich Cepheids are less luminous than metal-poor 
ones of the same pulsation period.

From the observational point of view the effect is not so obvious.
Whilst \citet{Rom08} found that metal-rich stars are fainter,
some studies found evidence of an opposite relation (\citealt{Kenni98};
\citealt{Udalski01}; \citealt{Ciar02}).

To solve this problem it is crucial to determine the metal 
content of individual Cepheids even for those located in
external galaxies. The spectroscopic observational technology 
has just reached the required level for the Magellanic Clouds.
But for Cepheids in more distant galaxies it is still beyond the 
possibilities. It would be very useful if the chemical composition 
could be determined -- or at least estimated -- without the 
need of spectroscopic observations, e.g., from photometry. 
The atmospheric iron abundance, [Fe/H], is a reliable indicator of 
stellar metallicity. The goal is to find a connection between the 
shape of the light curve and the metal content of Cepheids.

For RR Lyrae stars, \citet{Kovacs95} and \citet{Jurcsik96}
elaborated a method to determine the value of [Fe/H] from
the shape of the light curves.
They used the $R_{21}, R_{31}$ and $\phi_{31}$ Fourier parameters
(see Sect.~\ref{decomposition}) to calculate the photometric 
[Fe/H] ratio. Since RR~Lyrae type variables involved in the 
calibration are single mode, radially pulsating stars like Cepheids, 
existence of such relationship can be expected for Cepheids, 
as well. Indeed, \citet{Zsoldos95} found a similar relationship
based on a small sample of Cepheids due to the lack of available 
spectroscopic observations. Moreover the photometric data involved
were much less accurate than more recently.

In \citet[][hereafter Paper~I]{klagyi09}
we compiled a homogeneous data base of Galactic classical Cepheids. 
It contains the photometric and radial velocity peak-to-peak 
amplitudes, some amplitude related parameters, as well as information 
about the pulsation mode, binarity, and the spectroscopic 
[Fe/H] ratio. In \citet[][hereafter Paper~II]{szab12a} we investigated 
the metallicity dependence of the peak-to-peak pulsation amplitudes. 
The next step is the study of the influence of the [Fe/H] 
on the shape of the light curve.

In this paper we derive relationships between the $R_{21}$ and 
$R_{31}$ Fourier parameters and the spectroscopically determined 
atmospheric iron abundance, [Fe/H]. The Fourier decomposition of our 
Cepheid sample is presented in Sect.~\ref{decomposition}.
In Sect.~\ref{Fe/H-Four} the [Fe/H] dependences of the Fourier parameters 
are derived. In Sect.~\ref{phot_Fe} we apply these relationships for
determining the photometric [Fe/H] ratio of 14 Cepheids that have no 
spectroscopic iron abundance yet.
In Sect.~\ref{generality} we discuss the 
generalisation of the relations. Finally, Sect.~\ref{conclusion} 
contains our conclusions.

\section{Fourier decomposition}
\label{decomposition}

Since the light curves of the Cepheids are strictly periodic, 
they can be described with a sum of harmonic terms:
\begin{equation}
m(t) = A_0 + \sum_{i=1}^{N} a_i \cos(i\omega(t-t_0)) + 
\sum_{i=1}^{N} b_i \sin(i\omega(t-t_0))
%%%\label{equation~1}
\end{equation}
where $m(t)$ is the observed magnitude at time $t$, $A_0$ is the mean 
magnitude, $a_i$ and $b_i$ are the amplitudes of the $\sin$ and $\cos$
terms. The pulsation period of the Cepheid is $P = 2\pi / \omega$. 

Equation~1
can be written as
\begin{equation}
m(t) = A_0 + A_i \sum_{i=1}^{N}\cos(i\omega(t-t_0)+\phi_i)
\end{equation}
where $A_i = \sqrt{a_i^2 + b_i^2}$ and $\tan{\phi_i} = -b_i / a_i$.

For practical reasons \citet{SL81} introduced the dimensionless 
relative Fourier parameters that have been widely used ever since:

\begin{equation}
 R_{i1} = \frac{A_i}{A_1}~~{\rm and}~~\phi_{i1} = \phi_i - i\phi_1.
\end{equation}

The Fourier decomposition was made with the 
light curve characterization pipeline developed within Coordination Unit~7 
(Variability Processing) of the {\it Gaia} Data Processing and Analysis 
Consortium, with the procedure described in \citet{Detal11}.

In fact, the unprecedented precision of the photometry by {\it Kepler}
spacecraft has shown that the pulsation of V1154~Cygni, the only
Cepheid in the {\it Kepler} field is not strictly repetitive
\citep{Detal12}. Both the period and the shape of the light curve
are subjected to cycle-to-cycle variations but our approximation 
can be applied to the average light curve.

Classical Cepheids in the period range of $5 < P < 15$ days show 
a bump both in the photometric and radial velocity phase curves. 
This is the well known Hertzsprung progression \citep{Hertz26}.
This bump appears on the descending part of the light curve for
shorter periods, it is close to the maximum brightness around 10 
days and shifts toward even earlier phases for longer periods.
Due to this feature the shape of the light curve is
complex and an automated fitting method could not
retrieve the phase curve properly. To solve this problem the 
number of fitted harmonics ($N$) has been a function of the
pulsation period (see Table~\ref{num_harm}). Around 10 days 
consideration of more harmonics is necessary to fit the 
double peaked shape of the light curve than at much shorter or
longer pulsation periods. We did not use any period search algorithm, 
the pulsation periods were fixed during the Fourier
decomposition and have been taken from Paper~I.

\begin{table}
\caption{Number of fitted harmonics.}
\label{num_harm}
\centering
\begin{tabular}{cc}
\hline
Period (days) & $N$ \\
\hline
$P < 5.0$ & 3 \\
$5.0 \leq P < 8.0$ & 5 \\
$8.0 \leq P < 12.0$ & 9 \\
$12.0 \leq P < 15.0$ & 7 \\
$15.0 \leq P$ & 5 \\
\hline
\end{tabular}
\end{table}
\label{table-fit}

In almost every light curves there were some outlying 
observational data points. Since these outliers can significantly
modify the fit, we removed these points using a simple sigma clipping 
at $3 \sigma$ in the residual light curve after the first fit.
In the cleaned light curve -- due to the decreased scatter --
some other points can deviate by more than the new value
of $3 \sigma$ from the new fit. Therefore an iterating process
was applied until there were no more observed point beyond $3 \sigma$ 
from the fit. Two or three steps were usually sufficient. 
To have a satisfactorily large number of `useful' observational data, 
we set the limit of the fraction of points to ignore to 20\%.
To keep only the reliable fits, light curves covered with less than 
20 observed points have been omitted.

A portion of the final Fourier parameters are presented here
in Table \ref{tab:fourier} for guidance regarding its form and content.
Table \ref{tab:fourier} is published in its entirety in the electronic 
edition of this journal.

%%%%Table 2
\begin{table*}
 \label{tab:fourier}
 \caption{Fourier amplitude ratios of Galactic classical Cepheids. 
This table is published in its entirety in the electronic edition of this 
journal. The columns are as follows: name of the Cepheid,
pulsation period, spectroscopic [Fe/H] ratio, $R_{21}$ and $R_{31}$ 
in $B$, $V$, $R_C$ and $I_C$ bands, respectively and
information in binarity ($b$ denotes known binary/multiple system).}
\begin{center}
\begin{tabular*}{0.8\textwidth}{rcc|cc|cc|cc|cc|c}
\hline
Name & Period&[Fe/H]$_{\rm sp}$&\multicolumn{2}{|c|}{B} & \multicolumn{2}{|c|}{V} & 
\multicolumn{2}{|c|}{$R_C$} & \multicolumn{2}{|c|}{$I_C$} & Binarity \\
   & (d) & & $R_{21}$ & $R_{31}$ & $R_{21}$ & $R_{31}$ & $R_{21}$ & $R_{31}$ & $R_{21}$ & $R_{31}$ &  \\
\hline
\hline
   FM Aql & 6.114 &    0.29 & 0.361 & 0.103 & 0.360 & 0.120 & 0.354 & 0.105 & 0.365 & 0.152 &   \\
V1162 Aql & 5.376 &    0.06 & 0.308 & 0.074 & 0.335 & 0.086 & 0.339 & 0.099 & 0.348 & 0.120 &   \\
   EW Aur & 2.660 & $-$0.49 & 0.459 & 0.229 & 0.468 & 0.261 & 0.455 & 0.229 &  $-$  &  $-$  &   \\
   GV Aur & 5.260 & $-$0.16 & 0.461 & 0.208 & 0.450 & 0.194 & 0.455 & 0.205 &  $-$  &  $-$  &   \\
 V335 Aur & 3.413 & $-$0.25 & 0.439 & 0.210 & 0.453 & 0.244 & 0.460 & 0.222 &  $-$  &  $-$  &   \\
   AB Cam & 5.788 & $-$0.03 & 0.441 & 0.193 & 0.447 & 0.182 & 0.467 & 0.182 &  $-$  &  $-$  &   \\
   AC Cam & 4.157 & $-$0.08 & 0.427 & 0.110 & 0.412 & 0.145 & 0.410 & 0.149 &  $-$  &  $-$  &   \\
   RY CMa & 4.678 &    0.07 & 0.391 & 0.162 & 0.389 & 0.167 & 0.400 & 0.182 & 0.375 & 0.171 &  b\\
   RZ CMa & 4.255 &    0.02 &  $-$  &  $-$  & 0.369 & 0.127 &  $-$  &  $-$  & 0.375 & 0.111 &  b\\
   BC CMa & 4.175 &    $-$  &  $-$  &  $-$  & 0.364 & 0.213 &  $-$  &  $-$  & 0.327 & 0.184 &   \\
\hline
\end{tabular*}
\end{center}
\end{table*}

%%%%%%%%%%%%%%%%%%%%%%

\section{Metallicity dependence of the Fourier parameters}
\label{Fe/H-Four}

Owing to the increased interest in the metallicity dependence
of the $P$-$L$ relation, a number of recent papers deal 
with spectroscopic observations of Galactic -- and even extragalactic -- 
Cepheids (see the bibliographic list in Paper~II). In our sample
(taken from Paper~II), 329 out of 369 Cepheids have spectroscopic 
[Fe/H] values which means an impressive 90\% coverage.

There are various factors that have an influence on the observable
shape of the photometric light curve involved in determining the 
metallicity dependence. These factors include the pulsation period 
(short or long period, see Paper~I), the pulsation mode (fundamental 
or first overtone mode) and the binarity status. To have a reliable 
picture on the [Fe/H] dependence it is straightforward to sort the 
Cepheids according to their actual behaviour.

In Paper~I, we demonstrated that the short- and long-period Cepheids
behave in a different manner (Fig. 2. in Paper~I).
Therefore, it is worthwhile to separate the two groups.
The division between these two groups is at $\log{P}=1.02$ instead
of the period of 10 days adopted in most studies dealing with a 
break in the $P$-$L$ relationship (e.g., \citealt*{sandage09} and 
\citealt{ngeow09}). It should be noted that current theoretical
and empirical evidences indicate that the centre of the Hertzsprung
progression is metallicity dependent (\citealt{Bono00b} and references therein).
In this paper we only deal with the short-period 
Cepheids. The number of long period Cepheids in our sample is 
not large enough to deduce reliable relationships. 
Moreover the period dependence of the Fourier parameters is 
much stronger than in the case of short periods.

The pattern of the period dependence of the Fourier parameters 
of first overtone Cepheids is quite different from that of 
their fundamental mode counterparts, therefore overtone Cepheids 
cannot be treated together with the fundamental mode ones.
But their Galactic sample is not sufficiently large, either. 
Therefore in this paper we only deal with short period 
($\log{P} < 1.02$) and fundamental mode Cepheids.

	As to the binarity, a companion modifies the apparent
brightness, the peak-to-peak and Fourier amplitudes on magnitude scale
and usually the observed colors, too. 
But this is not the case in terms of Fourier amplitude ratios.
Using intensities instead of magnitudes, it is easy to prove that 
companions only contribute to the mean brightness but the amplitude 
ratios and phases remain unaffected. Therefore binaries can also be
included in our investigation. This is very useful for extragalactic 
Cepheids where photometric contamination due to blending is 
unavoidable. The importance of binary Cepheids is summarized by 
\citet{szab12b}.

\subsection{Period dependences}
\label{perioddependence}

 \begin{figure}
   \includegraphics[width=0.47\textwidth]{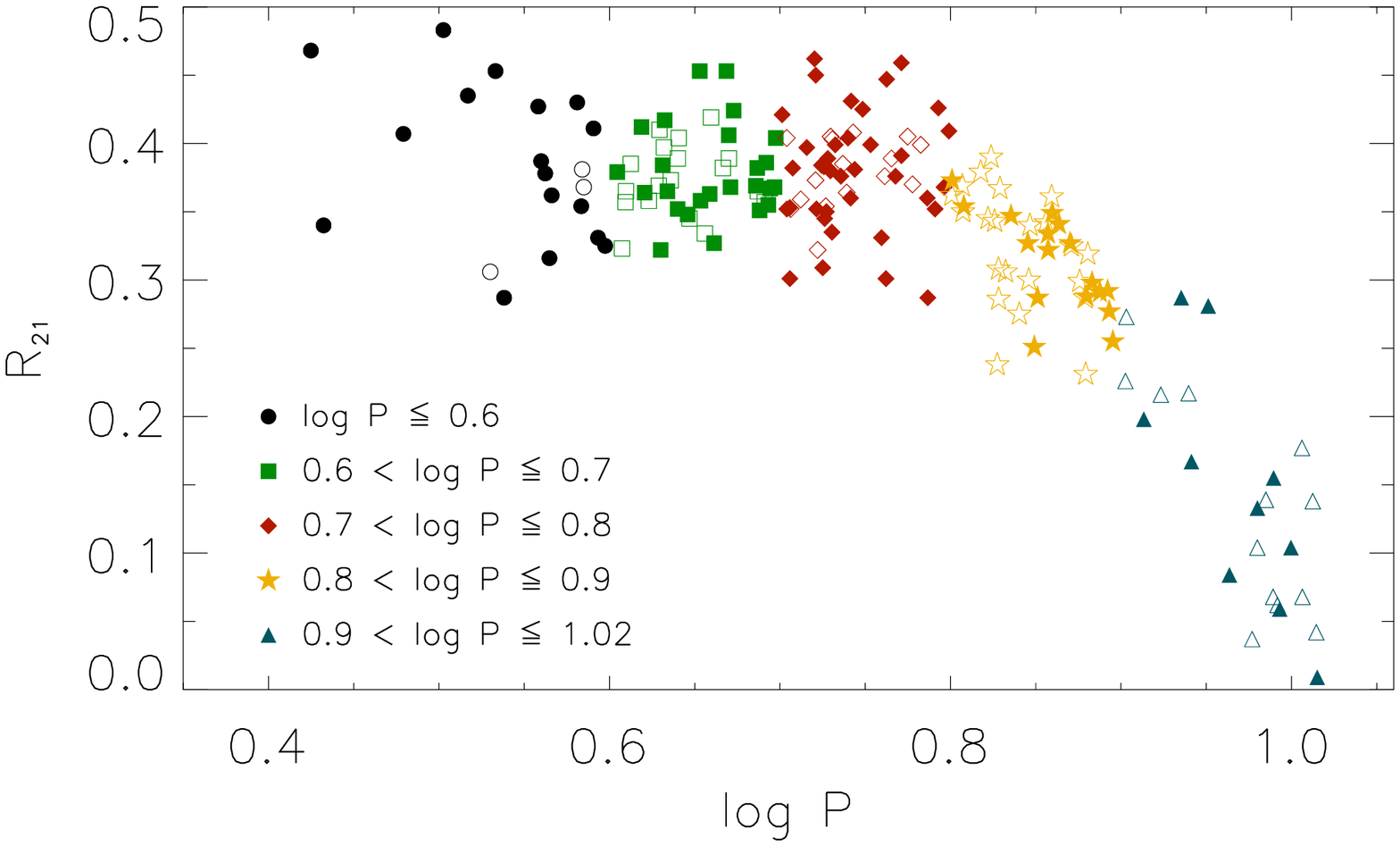}
   \includegraphics[width=0.47\textwidth]{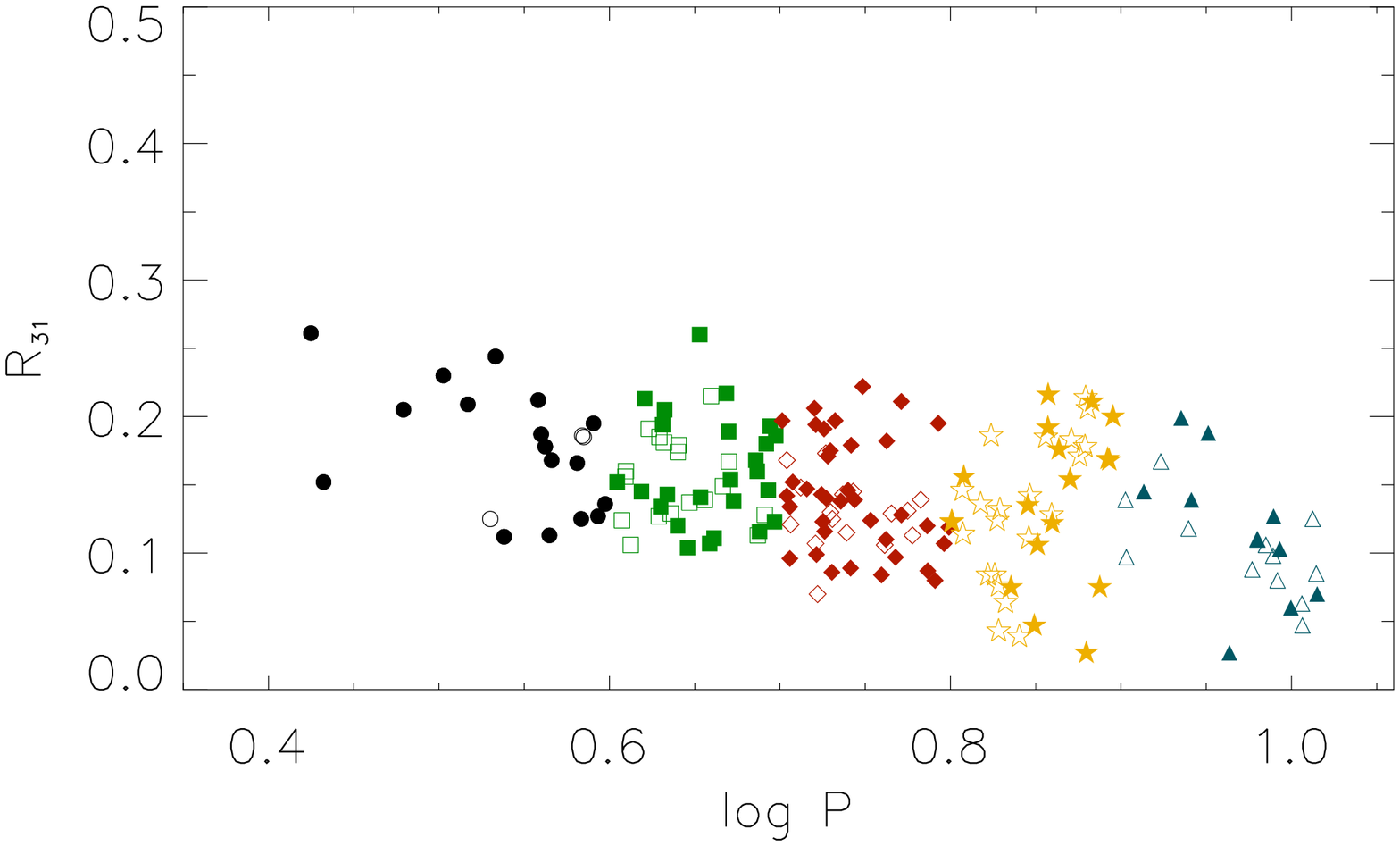}
   \caption{$V$ band $R_{21}$ and $R_{31}$ Fourier amplitude ratios 
   vs. $\log P$ for fundamental mode Cepheids (upper and lower
   panel, respectively).
%   The meaning of the symbols is as follows: 
%$\bullet$: $\log{P} \leq 0.6$; 
%$\blacksquare$: $0.6 < \log{P} \leq 0.7$; 
%$\blacklozenge$: $0.7 < \log{P} \leq 0.8$;
%$\bigstar$: $0.8 < \log{P} \leq 0.9$; 
%$\blacktriangle$: $0.9 < \log{P} \leq 1.02$. 
Filled symbols denote solitary Cepheids, while empty symbols represent 
known binaries.
}
   \label{r21v_logP}
 \end{figure}

To determine the metallicity dependence of the Fourier parameters
first we have to get rid of the effects influencing the
relationship to be determined. 
In the case of $R_{21}$ the most important is the pulsation period
(see the upper panel of Fig.~\ref{r21v_logP}).
The fall of $R_{21}$ at $\log{P} > 0.8$ in the upper panel 
of Fig.~\ref{r21v_logP} is a consequence of the 2:1 resonance between 
the fundamental and the second overtone modes of the radial pulsation 
\citep{andreasen87}.

In the case of $R_{31}$ the period dependence is smooth in 
the period interval $0.4 < \log P < 1.02$ (see the lower 
panel of Fig.~\ref{r21v_logP}).

The pattern of the period dependence of $R_{21}$ for short 
pulsation periods differs from the picture seen for the Cepheids both in the
Large and Small Magellanic Clouds (top left panel of Fig.~5 in 
\citealt[][]{Sosz08} and the upper panel of Fig.~2 
in \citealt{Sosz10}, respectively).
In both Magellanic Clouds this period domain of the figure 
can be described with an increasing and a decreasing part and with 
a local maximum at $\log{P} = 0.4$. It should be noted that our 
sample is much smaller than the Large and Small Magellanic Clouds
(hereinafter LMC and SMC, respectively) samples, both containing 
thousands of Cepheids and we lack Cepheids with $\log{P} < 0.4$ to 
reach a conclusion on the local maximum in our database, but
Galactic Cepheids do not show the decrease in $R_{21}$ for $\log{P} > 0.4$ (Fig. \ref{r21v_logP}).

Because our sample is not large enough to fit an accurate 
curve to the points in the top panel of Fig.~\ref{r21v_logP}, 
we grouped the Cepheids into several period intervals.
Each of these intervals ranges 0.1 in $\log{P}$ between
0.4 and 0.9, while for the longest periods the interval ranges [0.9,1.02].

\subsection{Fourier amplitude ratios vs. [Fe/H]}
\label{ampratios}

In contrast to the RR~Lyrae stars having a metallicity 
in the interval $-2.0 <$ [Fe/H] $< 0.0$, Galactic Cepheids 
are characterised with a much narrower interval of iron 
abundance.
The metallicity range is $-0.5 < {\rm [Fe/H]} < 0.5$
(\citealt{Pedicelli10} and \citealt{Luck11}). Our sample represents the
same range.
Moreover the most metal poor values belong to the faintest (i.e., 
most distant) stars, therefore the uncertainties are the largest.
The average uncertainty of the [Fe/H] value for 
a Cepheid is $\sim 0.1$ dex. Therefore, it is more difficult 
to detect any relationship between the [Fe/H] value and 
other properties of Cepheid variables.

To keep the data as homogeneous as possible we only used the 
multicolour photometric observations obtained by \citet{berd08} 
and his co-workers. 
Unfortunately the precision of these photometric data is not 
sufficiently good or in some cases the coverage of the phase curve 
is not satisfactory for obtaining reliable Fourier phases and then 
reveal any significant [Fe/H] dependence of these Fourier parameters. 
The Fourier amplitudes, however, are much less sensitive 
to the quality of the photometric observations and the phase 
coverage of the light curve.

   \begin{figure}
   \centering
   \includegraphics[width=0.47\textwidth]{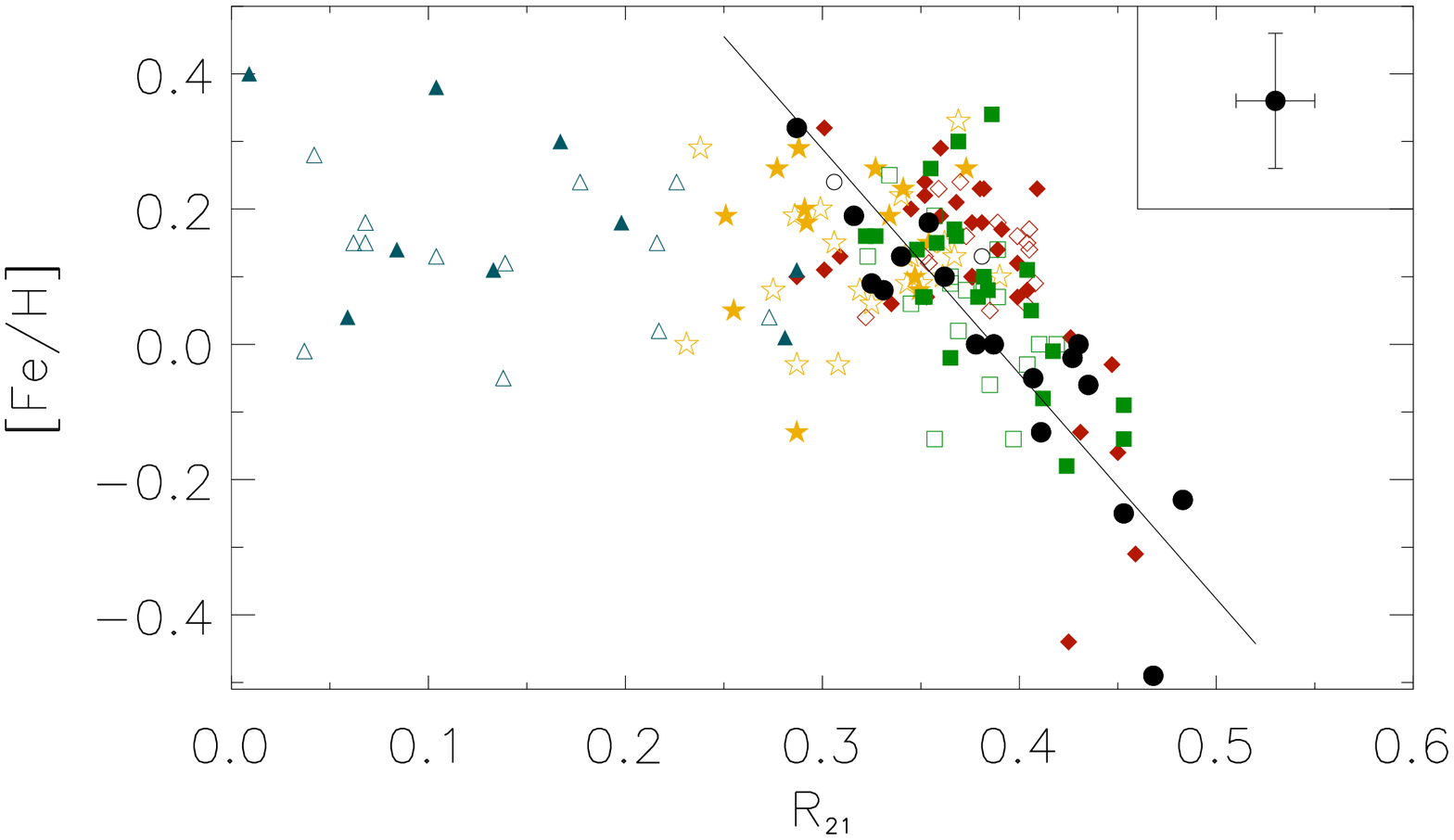}
   \includegraphics[width=0.47\textwidth]{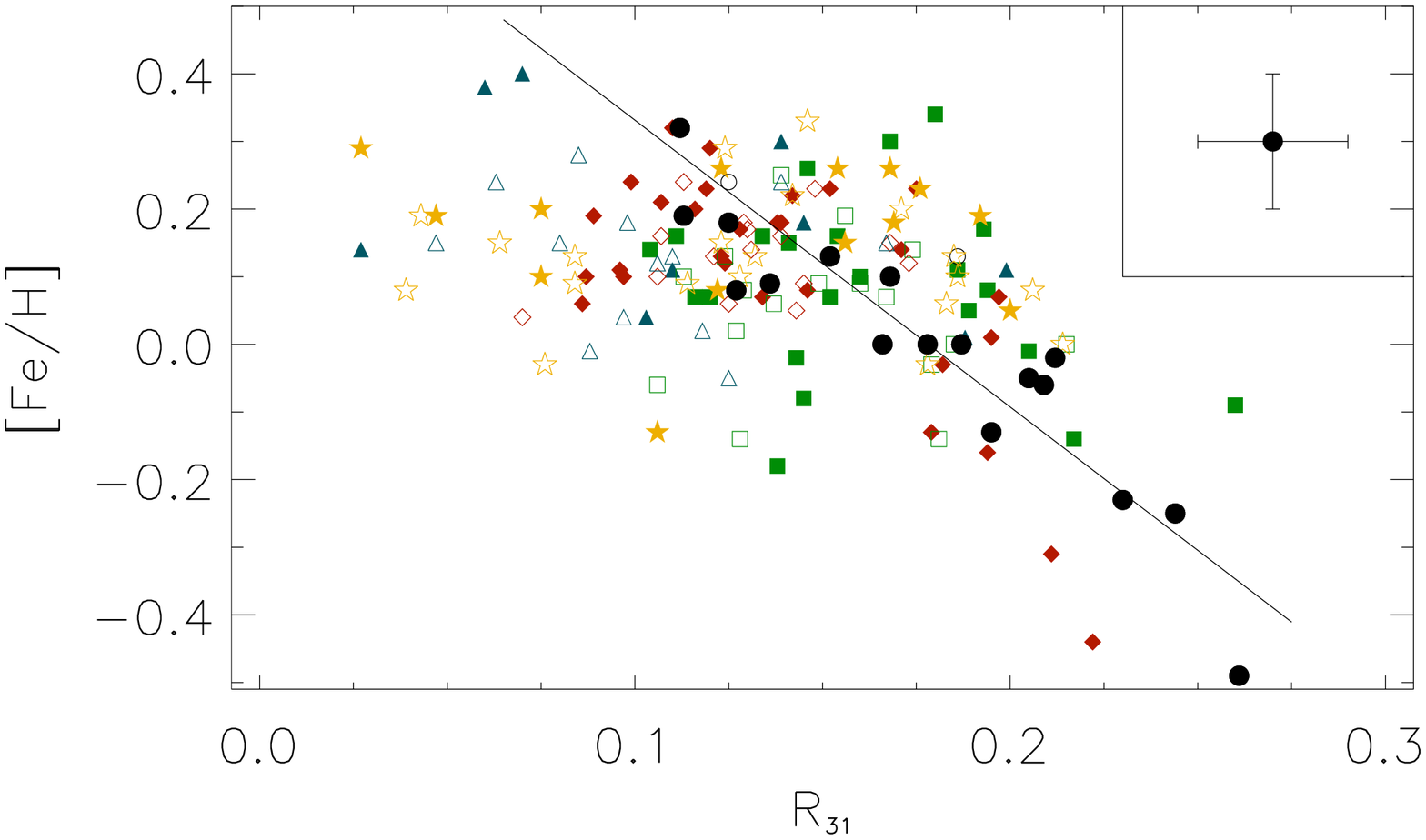}
   \caption{$V$ band $R_{21}$ and $R_{31}$ Fourier amplitude ratio vs. [Fe/H] 
   for fundamental mode Cepheids. The meaning of the symbols is the same as in Fig.~\ref{r21v_logP}. The fit
corresponds to the subsample with $\log{P} < 0.6$. Mean uncertainties are shown in the upper right insets.
   }
       \label{r21v_Fe}%
    \end{figure}

Figure~\ref{r21v_Fe} shows the metallicity dependence of the 
$R_{21}$ and $R_{31}$ amplitude ratios in $V$ band. 
For the shortest period Cepheids ($\log{P} < 0.6$) the
tendency is clearly seen: {\em the larger the metallicity,
the smaller the $R_{21}$ and $R_{31}$ amplitude ratios}.
Involving stars with longer pulsation periods (up to
$\log P = 0.8$) only the scatter around the ridge line fit
increases, the steepness does not change significantly.
The metallicity dependence of $R_{21}$ for $\log{P} > 0.8$
stars cannot be determined. In this period region
the Hertzsprung progression dominates.
This dominance does not affect the $R_{31}$ parameter in our Galactic sample,
however, a reliable fit can be achieved up to $\log P = 0.8$, too.
For longer period Cepheids ($\log P > 0.8$) there is no
correlation between the spectroscopic iron content and the $R_{31}$ parameter.

Figure~\ref{r21v_Fe} shows the results only in $V$ band, 
but the relationship seems to be independent of the
wavelength: the coefficients of the fits are the same within 
$1 \sigma$ in $B$, $V$ and $R_C$ photometric bands. 
The values of the coefficients are listed in Table~\ref{tab_fou_fe}.
The linear fit corresponds to the formula:

\begin{equation}
 {\rm [Fe/H]} = a + b \times R_{i1}
\label{fittingequation}
\end{equation}

\noindent where $i = 2$ or $3$. In Fig.~\ref{r21v_Fe} the fit represents the $\log{P} < 0.6$ period range.

%%%%%%%%%%TABLE 3
\begin{table*}
\centering
\caption{Fitted coefficients for the [Fe/H] dependence of the $R_{21}$ 
and $R_{31}$ Fourier parameters in various bands.
The meaning of the coefficients is defined in Eq.~\ref{fittingequation}. 
The last columns of $R_{21}$ and $R_{31}$ part
($\sigma_{O-C}$) of the table describes the standard deviation of the 
difference between the observed ([Fe/H]$_{\rm sp}$)
and calculated ([Fe/H]$_{\rm phot}$) iron contents.
}
\begin{tabular}{cccccccccccccccc}
\hline
\noalign{\smallskip}
 && \multicolumn{14}{c}{$\log{P} < 0.6$}\\
 && \multicolumn{6}{c}{$R_{21}$} && &\multicolumn{6}{c}{$R_{31}$}\\
\noalign{\smallskip}
\cline{3-8} \cline{11-16}
\noalign{\smallskip}
Band && a & $\sigma_a$ & b & $\sigma_b$ & $N$ & $\sigma_{O-C}$ &&& a & $\sigma_a$ & b & $\sigma_b$ & $N$ & $\sigma_{O-C}$ \\
\hline
$B$   && 1.452 & 0.178 & $-$3.918 & 0.524 & 17 & 0.081 &&& 0.749 & 0.120 & $-$4.623 & 0.780 & 17 & 0.114\\
$V$   && 1.287 & 0.173 & $-$3.326 & 0.480 & 19 & 0.091 &&& 0.755 & 0.079 & $-$4.240 & 0.458 & 19 & 0.082\\
$R_C$ && 1.314 & 0.258 & $-$3.535 & 0.707 & 13 & 0.113 &&& 0.702 & 0.142 & $-$4.262 & 0.893 & 13 & 0.116\\
$I_C$ && 0.859 & 0.071 & $-$2.182 & 0.178 & 10 & 0.055 &&& 0.538 & 0.108 & $-$2.852 & 0.613 & 10 & 0.090\\
\hline
\hline
\noalign{\smallskip}
 && \multicolumn{14}{c}{$\log{P} < 0.7$}\\
 && \multicolumn{6}{c}{$R_{21}$} && &\multicolumn{6}{c}{$R_{31}$}\\
\noalign{\smallskip}
\cline{3-8} \cline{11-16}
\noalign{\smallskip}
Band && a & $\sigma_a$ & b & $\sigma_b$ & $N$ & $\sigma_{O-C}$ &&& a & $\sigma_a$ & b & $\sigma_b$ & $N$ & $\sigma_{O-C}$ \\
\hline
$B$   && 1.363 & 0.177 & $-$3.512 & 0.487 & 44 & 0.099 &&& 0.585 & 0.117 & $-$3.474 & 0.802 & 44 & 0.132\\
$V$   && 1.372 & 0.137 & $-$3.479 & 0.363 & 58 & 0.103 &&& 0.594 & 0.079 & $-$3.333 & 0.473 & 58 & 0.132\\
$R_C$ && 1.214 & 0.203 & $-$3.122 & 0.553 & 41 & 0.111 &&& 0.514 & 0.121 & $-$2.884 & 0.787 & 41 & 0.133\\
$I_C$ && 0.984 & 0.096 & $-$2.582 & 0.256 & 30 & 0.099 &&& 0.501 & 0.073 & $-$2.900 & 0.424 & 30 & 0.124\\
\hline
\hline
\noalign{\smallskip}
 && \multicolumn{14}{c}{$\log{P} < 0.8$}\\
 && \multicolumn{6}{c}{$R_{21}$} && &\multicolumn{6}{c}{$R_{31}$}\\
\noalign{\smallskip}
\cline{3-8} \cline{11-16}
\noalign{\smallskip}
Band && a & $\sigma_a$ & b & $\sigma_b$ & $N$ & $\sigma_{O-C}$ &&& a & $\sigma_a$ & b & $\sigma_b$ & $N$ & $\sigma_{O-C}$ \\
\hline
$B$   && 1.379 & 0.167 & $-$3.499 & 0.456 & 76 & 0.123 &&& 0.569 & 0.078 & $-$3.417 & 0.549 & 76 & 0.134\\
$V$   && 1.375 & 0.131 & $-$3.426 & 0.346 &102 & 0.121 &&& 0.587 & 0.056 & $-$3.365 & 0.372 &102 & 0.128\\
$R_C$ && 1.269 & 0.185 & $-$3.170 & 0.500 & 76 & 0.131 &&& 0.534 & 0.074 & $-$3.067 & 0.530 & 76 & 0.134\\
$I_C$ && 0.986 & 0.104 & $-$2.485 & 0.285 & 53 & 0.123 &&& 0.353 & 0.067 & $-$1.813 & 0.496 & 53 & 0.124\\
\hline
\end{tabular}
\label{tab_fou_fe}
\end{table*}
%%%%%%%%%%

In the $I_C$ band the steepness of the fitted line is 
significantly smaller than for the other bands. 
The differences do not exceed $2 \sigma$ for the whole period 
domain investigated and for both the $R_{21}$ and $R_{31}$ amplitude ratios.
In order to compare the parameters in different bands correctly we fitted
the same function using only the stars that have $I$ band data.
The steepness of the slopes is closer to the $I$ band values
for each band but is still significantly lower.
This means that the brightness variation in the near infrared is less
sensitive to the atmospheric metallicity. It would be useful to
investigate this effect in the $J$, $H$ and $K$ infrared bands
but the amount of available observational data are not sufficient yet.

The absolute errors ($\sigma_a, \sigma_b$) of the fits for $R_{21}$ 
are systematically smaller in the $I_C$ band than using other filters. 
For $R_{31}$ the errors in $V$ band are comparable to the
$I_C$ band values and relative errors of the slope ($\sigma_b / b$) 
are smaller in the $V$ band for every period range.

There are some more methods that can be used to guess the
photometric [Fe/H] of Cepheids. \citet{Caputo01} adopted nonlinear
convective pulsational models to calculate the boundaries of
the $P$-$L$ relation for different chemical compositions. For observed Cepheids
the chemical composition can be guessed based on the position of the stars
within these boundaries. Unlike our one, this method can be used for all
period ranges and needs only the mean magnitude of the stars. But the
mean magnitude is affected by interstellar absorption and binarity.
The latter is less important for long period (i.e. bright) Cepheids,
but for short period stars, where our method is valid, it becomes
significant. Our method is not affected by binarity.

Another method was published by \citet{Pedicelli09}. They
calculated the photometric [Fe/H] ratio using multiband near infrared
($J$, $H$ and $K$) Walraven ($V$, $B$, $L$, $U$, $W$) photometry
complemented with evolutionary models. Since this method needs mean
magnitudes, too, it is also affected by the same effects.
The accuracy of this method is similar to that of our relations, $\sigma \sim 0.15$ dex.

The advantage of our method is its insensivity to many effects that can systematically
modify the calculated [Fe/H] ratios. But we are limited to a short range of
the pulsation periods.

\subsection{An outlook for extragalactic Cepheids}
\label{extragalactic}

To check whether the relations in Sect.~\ref{ampratios}
are valid for extragalactic Cepheids, too, we compared
the range of [Fe/H] and the Fourier amplitude ratios
of Cepheids in both the LMC and SMC.
However, we do not intend to validate our relations with the Magellanic Cloud Cepheids.

%%%%%%%%%Figure 3
   \begin{figure}
   \centering
   \includegraphics[width=0.47\textwidth]{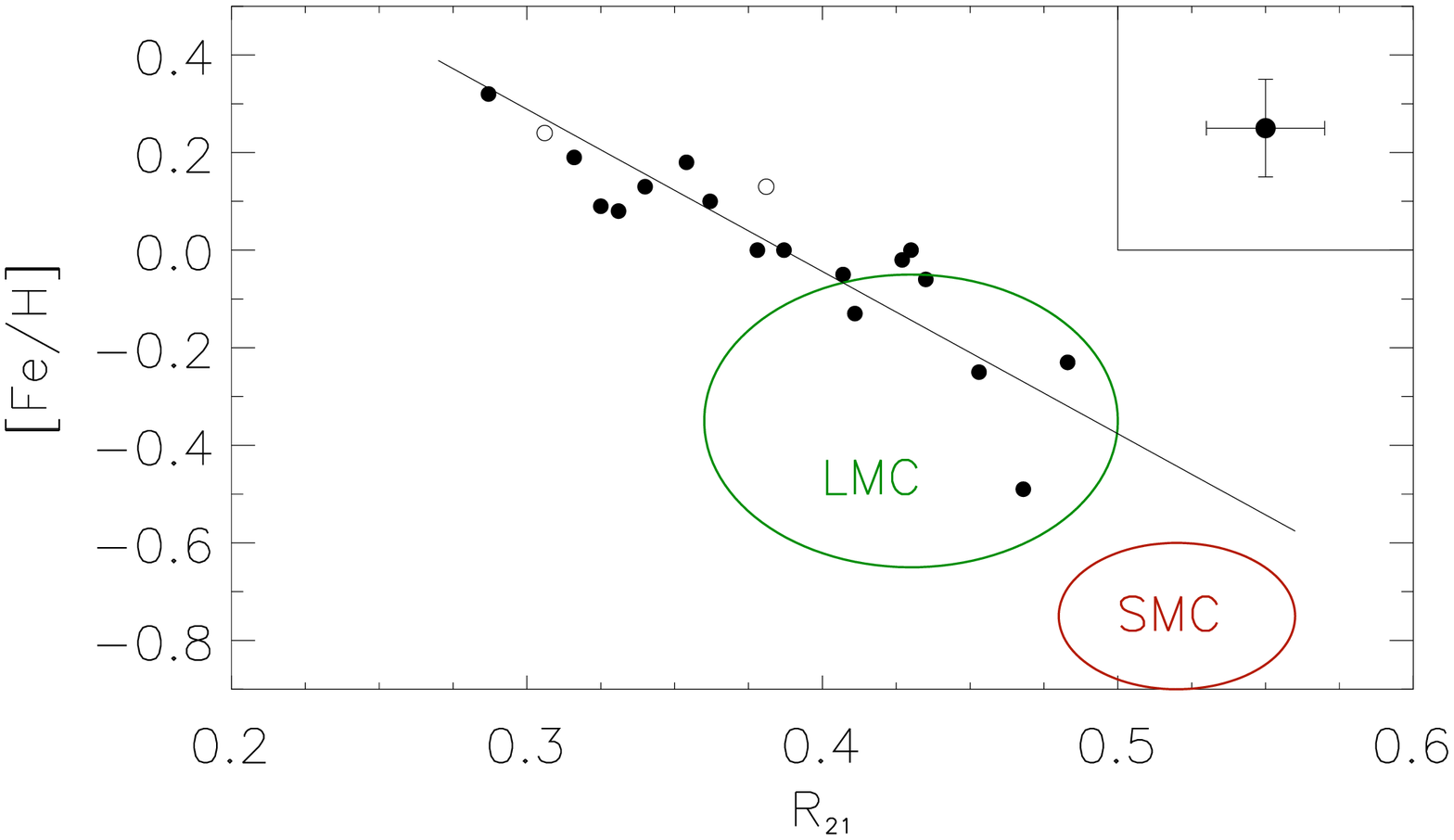}
   \includegraphics[width=0.47\textwidth]{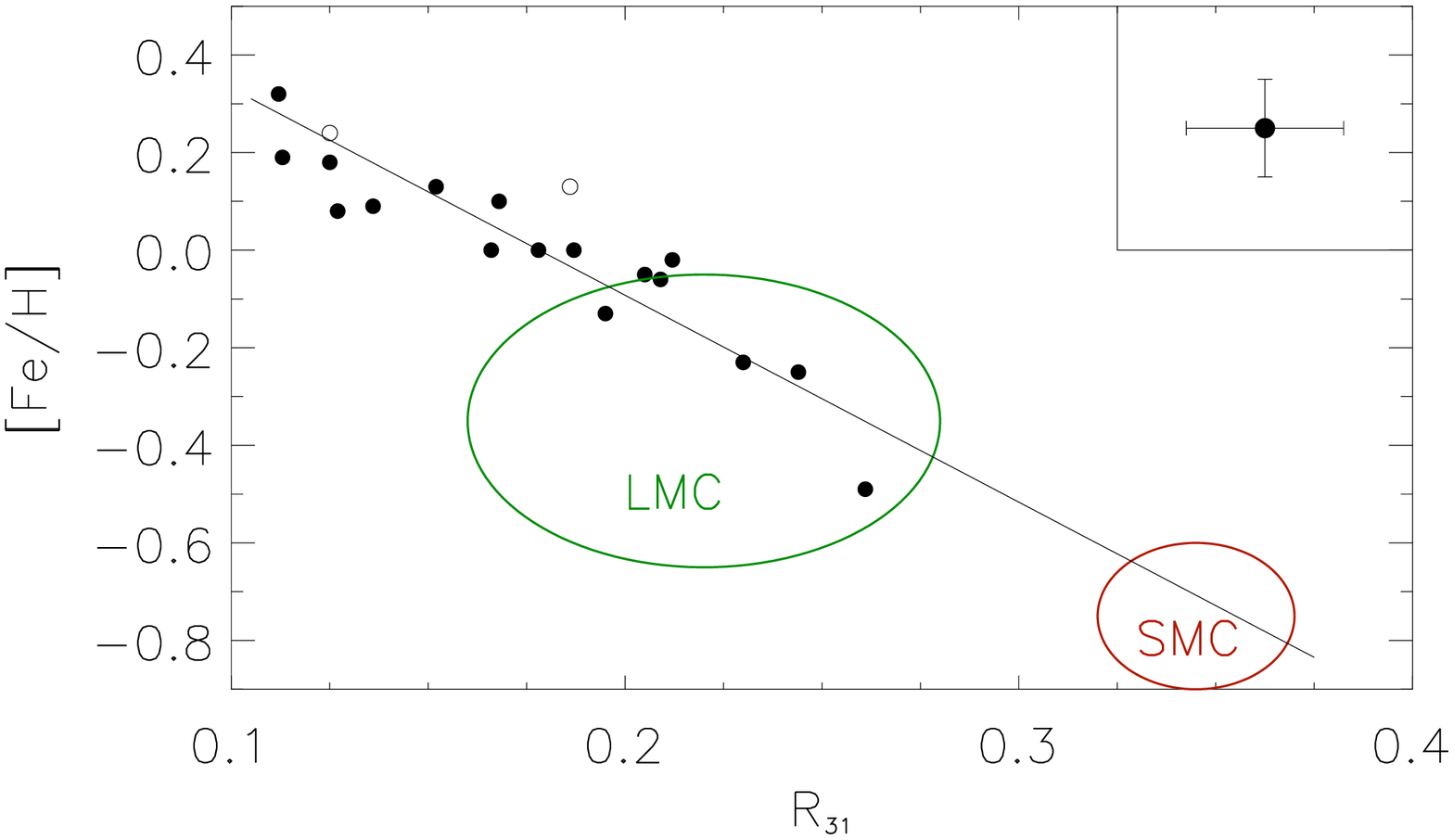}
   \caption{[Fe/H] vs. $V$ band $R_{21}$ and $R_{31}$ Fourier 
   amplitude ratios for fundamental mode Galactic Cepheids with 
   $\log{P} < 0.6$ and the range of the amplitude ratios the 
   Magellanic Cloud Cepheids for the same pulsation periods,
   details are in Sect.\ref{extragalactic}.
   The error bars in the upper right corner of both panels
   represent the average error of the Galactic Cepheids, 
   which is $\pm 0.02$ in $R_{21}$ and $R_{31}$ and $\pm 0.1$ 
   in [Fe/H]. The ellipses represent the range of the
   Magellanic Cloud Cepheids.
   }
    \label{r21v_Fe_magellanic}%
    \end{figure}

The pattern of the period dependence of the $R_{21}$ and $R_{31}$
Fourier parameters is slightly different for the Magellanic Clouds
and our Galaxy. For the LMC and SMC there is a peak at $\log{P} \sim 0.4$.
The corresponding $R_{21}$ values for the SMC are larger than for our 
Galactic sample, while for the LMC they are overlapping with the 
most metal-poor Galactic Cepheids. For the LMC $0.36 < R_{21} < 0.50$ 
and $0.16 < R_{31} < 0.28$ \citep{Sosz08}, while for the SMC 
$0.48 < R_{21} < 0.56$ and $0.32 < R_{31} < 0.37$ \citep{Sosz10} for most of the Cepheids.

In order to avoid any difficulty caused by the period dependence of the
Fourier amplitude ratios we used only the Cepheids with $0.4 < \log{P} < 0.6$.
In this range the amplitude ratios are roughly independent of the period.

Since there are no spectroscopic [Fe/H] values for Cepheids with $\log{P} < 0.6$ in the
Magellanic Clouds we cannot check our relations on individual stars. Instead, we
use the full ranges of the iron content of the LMC and SMC \citep{Rom08}
and the ranges of the $R_{21}$ and $R_{31}$ parameters for the period domain $\log{P} < 0.6$.
The shortest period Cepheid with individual spectroscopic [Fe/H] value
is HV 6093 (Romaniello et al. 2008) with log P = 0.68 and [Fe/H]$=-0.60$.
At this period the $R_{31}$ Fourier amplitude ratio starts falling due to the
Hertzsprung progression and our relations are not valid any more.
%Since the position of the Hertzsprung progression is metallicity
%dependent \citep{Bono00b} we do not see this decrease in our galactic sample.

%The average metallicities of Cepheids are $\rm{[Fe/H]} = -0.33 \pm 0.13$ 
%for the LMC and $\rm{[Fe/H]} = -0.75 \pm 0.08$ for the SMC 
%\citep{Rom08} but these iron abundances have been determined by
%averaging 22 and 14 individual values obtained spectroscopically, 
%for Cepheids in the LMC and the SMC, respectively.
%\citet{Rom08} observed long period and hence more luminous Cepheids. 
%Therefore the average [Fe/H] values and the maxima of the Fourier 
%amplitude ratios do not belong to the same stars.
%Nevertheless these average values for both Magellanic Clouds 
%fit our relationship as is seen in the top panel of 
%Fig.~\ref{r21v_Fe_magellanic}.

Figure~\ref{r21v_Fe_magellanic} testifies that the [Fe/H] vs. $R_{31}$ relationship
can be used for extragalactic Cepheids, too. There is no need
for spectroscopic observation to guess the [Fe/H] ratio of individual 
faint pulsators. The atmospheric iron content can be inferred from
an accurate optical light curve with good phase coverage.
The [Fe/H] vs. $R_{21}$ relationship suggests a non-linear
shape, which is not seen in the Galactic data itself. We need
spectroscopic [Fe/H] values of individual short period Cepheids
in the Magellanic Clouds to check the linearity of the relationship.

In a recent study \citet{majaess13} found that the peak-to-peak 
amplitudes of long period extragalactic Cepheids are significantly 
smaller at lower [Fe/H] ratio. This is just the opposite of the 
behaviour of the peak-to-peak amplitudes of short period Galactic
Cepheids \citep{szab12a} and the $R_{21}$ and $R_{31}$ Fourier 
parameters in this paper, also deduced for short period Cepheids.
These results are however not in conflict with the theoretical 
calculations. The model computed by \citet{Bono00a} indicates that 
metal-rich Cepheids with a pulsation period of 10-30 days pulsate 
with larger peak-to-peak amplitudes than metal-poor ones.
A similar study would be needed for short period extragalactic
Cepheids to be sure that their observed behaviour conforms to
the theoretical calculations.

The short and long period Cepheids can have different iron content.
In our Galactic sample there is only a slight difference in the [Fe/H] ratio of the short and
long period Cepheids. Their average values are 0.096 ($\sigma = 0.137$) 
and 0.174 ($\sigma = 0.160$), respectively. Since the age difference
of short and long period Cepheids is small, this difference
might be a selection effect. Long period Cepheids are brighter
and therefore they can be detected at larger distances.
Due to the metallicity gradient of the Galaxy \citep{Pedicelli09}
they are more metal-rich on average if they are located toward the
Galactic center and they are more metal-poor if they are located outwards.
In our database we have more long period Cepheids toward the Galactic center.

Another difference is that the distribution of the pulsation periods is metallicity dependent
\citep{Bono00a}. The minimum mass evolutionary tracks crossing the
instability strip is decreasing with lower metallicity. Moreover,
the lifetime spent inside the instability strip is longer for
lower mass Cepheids. These result in a higher relative number of
shorter period Cepheids in the Magellanic Clouds.

Observing individual extragalactic Cepheids is hindered by the
fact that these stars are located in a crowded field of view 
and the spatial resolution of the telescopes is limited. Therefore 
the determination of accurate peak-to-peak amplitudes is hardly 
possible.
%However the Fourier amplitude ratios are not or only 
%slightly affected by the contribution from additional lights.

%%%%%%%%%%%%%%%%%%%%%%%%%%%%%%

\section{Photometric iron content}
\label{phot_Fe}

Our main goal was to elaborate a method that needs no 
spectroscopic observations to determine the metallicity of 
individual Cepheids. The relationships deduced above
between the photometric Fourier amplitude ratios and the 
spectroscopic iron content offer a viable solution.

%%%%%%%%

   \begin{figure*}
   \centering
   \includegraphics[width=0.856\textwidth]{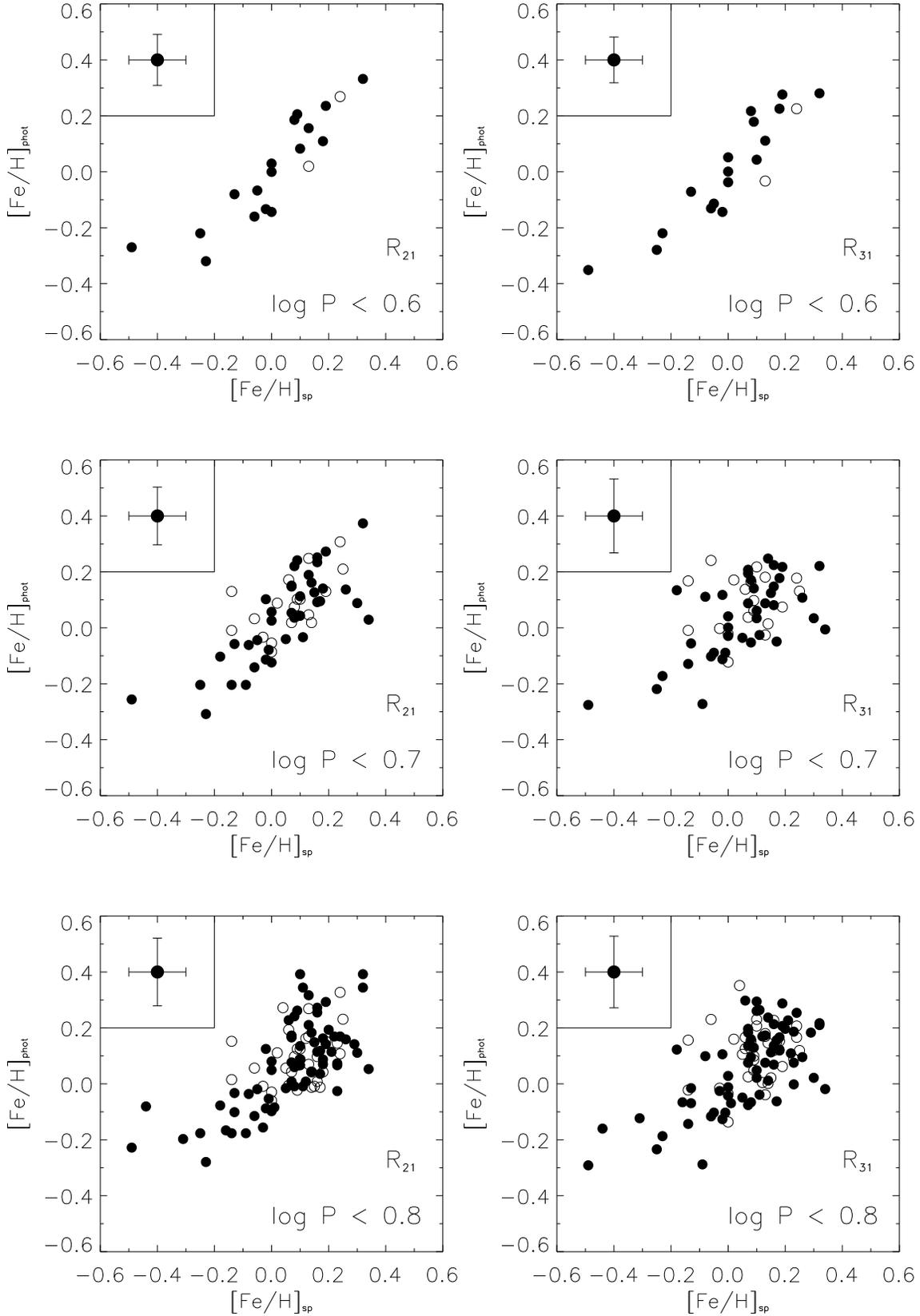}
   \caption{Spectroscopic [Fe/H] ratios vs. the calculated 
   photometric [Fe/H] ratios in the $V$ band for the relations in 
   Table~\ref{tab_fou_fe}. Filled symbols refer to solitary Cepheids, while
   binary Cepheids are marked with open circles. In the upper left corners
   we show the average error bars for each panel.
   }
              \label{obs_calc_Fe_1}%
    \end{figure*}

Based on the relations in Table~\ref{tab_fou_fe} we calculated the 
photometric [Fe/H] ratios for the suitable Cepheids.
The scatter of the difference between the observed and calculated 
[Fe/H] values were determined in order to guess the accuracy of the 
photometric metallicities -- see the last columns of the $R_{21}$ and 
$R_{31}$ part of Table~\ref{tab_fou_fe} and Fig.~\ref{obs_calc_Fe_1}). 
We only used the fundamental mode Cepheids both with and without known 
companion(s), but in the figures we plotted them with different symbols.
The precision of the photometric [Fe/H] values are significantly higher
in the $I_C$ band than in the others. In the $B$, $V$ and $R_C$ bands 
the scatter is nearly identical with a slight decrease for 
the $V$ band. The accuracy of the photometric [Fe/H] is 
$\sim 0.1$ dex when determined from $R_{21}$ for Cepheids with 
$\log{P} < 0.7$ or based on $R_{31}$ for Cepheids with $\log{P} < 0.6$.  
In the case of longer period Cepheids, the accuracy is $\sim 0.15$ dex.

The higher accuracy of the $I$ band fits was investigated.
Such an effect can be caused for example by significantly more precise photometric
observations or more precise Fourier amplitude determination of large
amplitude Cepheids. But the quality of the photometry does not differ too much
in different bands and we found that our sample represents the entire amplitude
range. These effects cannot play a role in the mystery of the accuracy.
However, if we calculate the scatter of the difference between
the observed and calculated [Fe/H] values using only the stars with $I$ band
data, the $V$ band standard deviation decreases from $0.091$ to $0.072$ for $R_{21}$
and $\log{P} < 0.6$. This is still higher than $0.055$ for the $I$ band.
So a part of the difference can be explained with the small number of datapoints
but the higher accuracy values for the $I$ band could be realistic.

%%%%%%%%%%
\begin{table*}
\caption{
Calculated $V$ band $\rm{[Fe/H]_{phot}}$ values for Cepheids without 
spectroscopic [Fe/H] ratio. The data in columns 3-10 refer to the period 
ranges mentioned at the top of each column.
In the columns 11-13 the average values are presented (average of columns 
3-5, 6-10, and 3-10, respectively).
The last column contains information on binarity ($b$: Cepheid with known 
companion(s)).}
\label{tab_phot_iron}
\begin{center}
\begin{tabular*}{0.85\textwidth}{rc|ccc|ccc|ccc|c}
\hline
\hline
 Cepheid& $\log{P}$ & \multicolumn{9}{|c|}{$\rm{[Fe/H]_{phot}}$ in $V$ band} & \\
 &  & \multicolumn{3}{|c|}{$R_{21}$} & \multicolumn{3}{|c|}{$R_{31}$} & \multicolumn{3}{|c|}{average} & \\
 & & $< 0.6$ & $< 0.7$ & $< 0.8$ & $< 0.6$ & $< 0.7$ & $< 0.8$ & $R_{21}$ & $R_{31}$ & all & bin.\\
\hline
   BC CMa & 0.621 &    -- &  0.11 &  0.13 &    -- &$-$0.12&$-$0.13&  0.12 &$-$0.12&   0.0 & \\
   CQ Car & 0.726 &    -- &    -- &  0.06 &    -- &    -- &$-$0.06&  0.06 &$-$0.06&   0.0 & \\
   IM Car & 0.727 &    -- &    -- &  0.18 &    -- &    -- &  0.12 &  0.18 &  0.12 &  0.15 & \\
   LV Cen & 0.697 &    -- &  0.09 &  0.11 &    -- &  0.18 &  0.17 &  0.10 &  0.18 &  0.14 & \\
   KO Cep & 0.659 &    -- &  0.11 &  0.13 &    -- &  0.24 &  0.24 &  0.12 &  0.24 &  0.18 & \\
   SU Cyg & 0.585 &  0.06 &  0.09 &  0.11 &$-$0.03&$-$0.02&$-$0.04&  0.09 &$-$0.03&  0.03 & b\\
   WY Pup & 0.720 &    -- &    -- &$-$0.21&    -- &    -- &$-$0.11&$-$0.21&$-$0.11&$-$0.16& \\
   WZ Pup & 0.701 &    -- &    -- &$-$0.07&    -- &    -- &$-$0.08&$-$0.07&$-$0.08&$-$0.07& \\
 V773 Sgr & 0.760 &    -- &    -- &  0.24 &    -- &    -- &  0.30 &  0.24 &  0.30 &  0.27 & \\
V1954 Sgr & 0.791 &    -- &    -- &  0.17 &    -- &    -- &  0.32 &  0.17 &  0.32 &  0.25 & \\
    X Sct & 0.623 &    -- &  0.13 &  0.15 &    -- &$-$0.04&$-$0.06&  0.14 &$-$0.05&  0.05 & b\\
   CR Ser & 0.724 &    -- &    -- &  0.06 &    -- &    -- &  0.11 &  0.06 &  0.11 &  0.09 & \\
   DP Vel & 0.739 &    -- &    -- &  0.13 &    -- &    -- &  0.20 &  0.13 &  0.20 &  0.17 & b\\
   BR Vul & 0.716 &    -- &    -- &  0.02 &    -- &    -- &  0.09 &  0.02 &  0.09 &  0.06 & \\
\hline
\end{tabular*}
\end{center}
\end{table*}
%%%%%%%%%%

There are 14 Cepheids in our sample that have no spectroscopic
[Fe/H] value yet, nevertheless we are able to calculate the photometric 
ones. For stars with the shortest periods the equations in 
Table~\ref{tab_fou_fe} result 24 different individual [Fe/H] values. 
Since the accuracy of the photometric iron content
in $V$ band is similar to that in the $I_C$ band and all of the stars 
were observed in the $V$, in Table \ref{tab_phot_iron} we present the 
calculated values for the $V$ band. Based on Tables~\ref{tab:fourier} 
and \ref{tab_fou_fe} one can easily calculate all other [Fe/H] ratios.
In Table~\ref{tab_phot_iron} we also calculated the average 
of the $R_{21}$-based, the $R_{31}$-based iron contents and the average 
of all the individual $V$ band values. The standard deviation
of the $observed - calculated$ values of these averages are $0.118$, 
$0.133$, and $0.123$, respectively, based on the Cepheids with known [Fe/H]$_{\rm sp}$.

The photometric [Fe/H] ratios based on $R_{21}$ and $R_{31}$ are similar
to each other within $\sim0.1$ dex in most cases. The two exceptional 
cases are BC~CMa and X~Sct where the difference is $\sim0.2$.

\section{Discussion}
\label{generality}

In Sect.~\ref{ampratios} we showed that practically identical 
relations are valid for the $B$, $V$ and $R_C$ photometric bands 
and a similar relation was determined for the $I_C$ band.
Since the {\it Gaia} space probe will also observe in the optical 
range, a careful calibration will facilitate using {\it Gaia} 
photometry for determining the iron content of thousands of target 
Cepheids independently of the spectroscopically derived [Fe/H] value
obtained from the measurements during this astrometric space project.
Since the companions in binary or multiple systems have no 
effect for our study, the photometric [Fe/H] ratio 
can be calculated for such Cepheids, too.

The {\it Gaia} astrometric space probe will scan the whole sky 
down to 20th magnitude and will provide a complete and unbiased data 
base involving an average of 80 photometric points on each target 
during its 5 year active lifetime. For Cepheids pulsating with 
a single period and stable light curve these $\sim$80 photometric data 
points seem to be sufficient to obtain a well covered phase curve over 
the complete pulsation cycle. To calculate precise Fourier amplitude 
ratios, a photometric accuracy of $\sim 0.02$ magnitude is sufficient. 
In this case, the photometric metallicity of Cepheids can be derived
down to the brightness range of 16-18th magnitudes \citep{Jordi10} 
depending on the photometric band and the colour of the target star,
which is much beyond the limit of ground based spectroscopic
observations yet.

Our method can be applied for extragalactic Cepheids, too.
The applicability of our formulae was shown for Cepheids 
in the Magellanic Clouds (see Sect.~\ref{extragalactic}).
Since Cepheids are brighter in the near infrared region than
at optical wavelengths, photometric observations of extragalactic 
Cepheids overwhelm in the near infrared, especially in the $I_C$ band.
Since the relationship between the atmospheric iron content and the
light curve shape derived by us (see Table~\ref{tab_fou_fe}) is also 
reliable for the photometric $I_C$ band, our formula offers a promising 
method for determination of metallicity of faint extragalactic Cepheids 
from purely photometric data.

The most accurate determination of the photometric [Fe/H] value 
can be performed for Cepheids with a pulsation period shorter than 
4~days. Since the absolute visual magnitude of these Cepheids is around 
$M_V = -2$ and the distance modulus of the LMC and SMC are $18.45$ mag 
and $18.92$ mag, respectively \citep{Storm11}, these stars have an apparent 
magnitude $m_V\sim16-17$, which is within the capability of the {\it Gaia}
photometry.

The photometric [Fe/H] value for a huge number of Galactic and
extragalactic Cepheids together with the extremely accurate 
trigonometric parallaxes of these stars to be provided by 
{\it Gaia} can help to recalibrate the metallicity dependence 
of the $P$-$L$ relationship and to refine the cosmic distance scale.

\section{Conclusions}
\label{conclusion}

We have presented a method to estimate the atmospheric iron content
of classical Cepheids using only photometric observations.
Based on a homogeneous data set of 369 Galactic Cepheids we
determined the relationships between the $R_{21}$ and $R_{31}$ 
Fourier amplitude ratios and the spectroscopic [Fe/H] ratio.

We found that for both $R_{21}$ and $R_{31}$ parameters the
higher the metal content, the smaller the amplitude ratio.
These relationships are valid and are the same for the 
Johnson $B$ and $V$ and the Kron-Cousins $R_C$ photometric bands.
For the $I_C$ band a similar but shallower relationship
is valid with an even better accuracy as for $V$ band.

The average [Fe/H] ratios and the range of the $R_{21}$ and 
$R_{31}$ amplitude ratios for the Magellanic Clouds are in 
good agreement with our relations, especially for $R_{31}$.
So the equations may be used also for at least the LMC and SMC.

We calculated the photometric [Fe/H] ratio for the suitable Cepheids.
14 of them have no spectroscopic iron content yet.
The accuracy of these values is $\sim 0.1$ dex for Cepheids pulsating
with the shortest period which is similar to
the error of the spectroscopic [Fe/H] value. For longer periods
the accuracy is $\sim 0.15$ dex. The relations are valid for
the pulsation periods up to $\log{P} = 0.8$.

Since binarity does not modify the Fourier amplitude 
ratios, therefore these relationships can be used to estimate 
the atmospheric iron content for binary Cepheids or for Cepheids
in crowded fields (e.g. for extragalactic Cepheids), too.

\section*{Acknowledgments}

Financial support from the ESA PECS Project C98090,
ESTEC Contract No.4000106398/12/NL/KML and support from the
Hungarian State E\"otv\"os Fellowship is gratefully acknowledged.
The authors are indebted to the referee, Dr. Giuseppe Bono, whose
critical report helped improve the presentation of the results.

\label{lastpage}


\begin{thebibliography}{99}
   \bibitem[\protect\citeauthoryear{Andreasen \& Petersen}{1987}]{andreasen87} Andreasen G.~K., Petersen J.~O., 1987, A\&A, 180, 129

   \bibitem[\protect\citeauthoryear{Baraffe \& Alibert}{2001}]{Bara01} Baraffe I., Alibert Y., 2001, A\&A, 371, 592

   \bibitem[\protect\citeauthoryear{Berdnikov}{2008}]{berd08} Berdnikov L.~N., 2008, http://vizier.u-strasbg.fr/viz-bin/VizieR?-source=II/285

   \bibitem[\protect\citeauthoryear{Bono et~al.}{1999}]{Bono99} Bono G., Caputo F., Castellani V., Marconi M., 1999, ApJ, 512, 711
   
   \bibitem[\protect\citeauthoryear{Bono, Castellani \& Marconi}{Bono~et~al.}{2000a}]{Bono00a}Bono G., Castellani V., Marconi M., 2000a, ApJ, 529, 293
   
   \bibitem[\protect\citeauthoryear{Bono, Marconi \& Stellingwerf}{Bono~et~al.}{2000b}]{Bono00b}Bono G., Marconi M., Stellingwerf R. F., 2000b, A\&A, 360, 245
   
   \bibitem[\protect\citeauthoryear{Caputo et~al.}{2000}]{Caputo00} Caputo F., Marconi, M., Musella I., Santolamazza P., 2000, A\&A, 359, 1059
   
   \bibitem[\protect\citeauthoryear{Caputo et~al.}{2001}]{Caputo01} Caputo F., Marconi M., Musella I., Pont F., 2001, A\&A, 372, 544
   
   \bibitem[\protect\citeauthoryear{Ciardullo et~al.}{2002}]{Ciar02} Ciardullo R., Feldmeier J.~J., Jacoby G.~H., Kuzio de Naray R., Laychak M.~B., Durrell P.~R., 2002, ApJ, 577, 31
   
   \bibitem[\protect\citeauthoryear{Derekas et~al.}{2012}]{Detal12} Derekas, A., et~al. 2012, MNRAS, 425, 1312

   \bibitem[\protect\citeauthoryear{Dubath et~al.}{2011}]{Detal11} Dubath P. et~al. 2011, MNRAS, 414, 2602

   \bibitem[\protect\citeauthoryear{Hertzsprung}{1926}]{Hertz26} Hertzsprung E., 1926, Bull. astronom. Inst. Netherl. 3, 115

   \bibitem[\protect\citeauthoryear{Jordi et~al.}{2010}]{Jordi10} Jordi, C., et~al. 2010, A\&A, 523, 48

   \bibitem[\protect\citeauthoryear{Jurcsik \& Kov\'acs}{1996}]{Jurcsik96} Jurcsik, J., Kov\'acs, G. 1996, A\&A, 312, 11

   \bibitem[\protect\citeauthoryear{Kennicutt et~al.}{1998}]{Kenni98} Kennicutt, R.~C., Jr., et~al., 1998, ApJ, 498, 181

   \bibitem[\protect\citeauthoryear{Klagyivik \& Szabados}{2009}]{klagyi09} Klagyivik, P., Szabados, L., 2009, A\&A, 504, 959 (Paper~I)

   \bibitem[\protect\citeauthoryear{Kov\'acs \& Zsoldos}{1995}]{Kovacs95} Kov\'acs, G., Zsoldos, E., 1995, A\&A, 293, 55L
    
   \bibitem[\protect\citeauthoryear{Luck \& Lambert}{2011}]{Luck11} Luck R. E., Lambert D. L., 2011, AJ, 142, 136
    
   \bibitem[\protect\citeauthoryear{Majaess et~al.}{2013}]{majaess13} Majaess, D., Turner, D.~G., Gieren, W., Berdnikov, L., Lane, D.~J., 2013, Ap\&SS, 344, 381

   \bibitem[\protect\citeauthoryear{Ngeow et~al.}{2009}]{ngeow09} Ngeow, C.-C., Kanbur, S.~M., Neilson, H.~R., Nanthakumar, A., Buonaccorsi, J., 2009, ApJ, 693, 691

   \bibitem[\protect\citeauthoryear{Pedicelli et~al.}{2009}]{Pedicelli09} Pedicelli S., et~al., 2009, A\&A, 504, 81

   \bibitem[\protect\citeauthoryear{Pedicelli et~al.}{2010}]{Pedicelli10} Pedicelli S., et~al., 2010, A\&A, 518, 11

   \bibitem[\protect\citeauthoryear{Romaniello et~al.}{2008}]{Rom08} Romaniello, M., et~al., 2008, A\&A, 488, 731

   \bibitem[\protect\citeauthoryear{Sandage, Bell \& Tripicco}{Sandage et~al.}{1999}]{sandage99} Sandage, A., Bell, R.~A., Tripicco, M.~J., 1999, ApJ, 522, 250

   \bibitem[\protect\citeauthoryear{Sandage, Tammann \& Reindl}{Sandage et~al.}{2009}]{sandage09} Sandage, A., Tammann, G. A., Reindl, B., 2009, A\&A, 493, 471

   \bibitem[\protect\citeauthoryear{Simon \& Lee}{1981}]{SL81}Simon, N.~R., Lee, A.~S., 1981, ApJ, 248, 291

   \bibitem[\protect\citeauthoryear{Soszy\'nski et~al.}{2008}]{Sosz08} Soszy\'nski, I., et~al., 2008, AcA, 58, 163

   \bibitem[\protect\citeauthoryear{Soszy\'nski et~al.}{2010}]{Sosz10} Soszy\'nski, I., et~al., 2010, AcA, 60, 17

   \bibitem[\protect\citeauthoryear{Storm et~al.}{2011}]{Storm11} Storm, J., Gieren, W., Fouqu\'e, P., Barnes, T.~G., Soszy\'nski, I., Pietrzy\'nski, G., Nardetto, N., Queloz, D., 2011, A\&A, 534, 95

   \bibitem[\protect\citeauthoryear{Szabados \& Klagyivik}{2012a}]{szab12a} Szabados, L., Klagyivik, P., 2012a, A\&A, 537, 81 (Paper~II)

   \bibitem[\protect\citeauthoryear{Szabados \& Klagyivik}{2012b}]{szab12b} Szabados, L., Klagyivik, P., 2012b, Ap\&SS, 341, 99

   \bibitem[\protect\citeauthoryear{Udalski et~al.}{2001}]{Udalski01} Udalski, A., Wyrzykowski, L., Pietrzynski, G., Szewczyk, O., Szymanski, M., Kubiak, M., Soszy\'nski, I., Zebrun, K., 2001, AcA, 51, 221

   \bibitem[\protect\citeauthoryear{Zsoldos}{1995}]{Zsoldos95} Zsoldos, E., 1995, in Astrophysical Applications of Stellar Pulsation, ed. R.~S. Stobie \& P.~A. Whitelock, ASPC, 83, 351
\end{thebibliography}
\end{document}